\documentclass[a4paper,10pt]{article}
\usepackage{graphicx}
\usepackage{comment}
\usepackage{amssymb}
\usepackage{amsmath}
\usepackage{nicefrac}
\usepackage{graphicx}
\usepackage{dcolumn}
\usepackage{bm}
\usepackage{comment}
\usepackage{multirow}
\usepackage{cite}
\usepackage{xcolor}

\begin{document}

\newcommand{\bin}[2]{\left(\begin{array}{c}\!#1\!\\\!#2\!\end{array}\right)}
\newcommand{\threej}[6]{\left(\begin{array}{ccc}#1 & #2 & #3 \\ #4 & #5 & #6 \end{array}\right)}
\newcommand{\sixj}[6]{\left\{\begin{array}{ccc}#1 & #2 & #3 \\ #4 & #5 & #6 \end{array}\right\}}
\newcommand{\regge}[9]{\left[\begin{array}{ccc}#1 & #2 & #3 \\ #4 & #5 & #6 \\ #7 & #8 & #9 \end{array}\right]}
\newcommand{\La}[6]{\left[\begin{array}{ccc}#1 & #2 & #3 \\ #4 & #5 & #6 \end{array}\right]}
\newcommand{\hj}{\hat{J}}
\newcommand{\hux}{\hat{J}_{1x}}
\newcommand{\hdx}{\hat{J}_{2x}}
\newcommand{\huy}{\hat{J}_{1y}}
\newcommand{\hdy}{\hat{J}_{2y}}
\newcommand{\huz}{\hat{J}_{1z}}
\newcommand{\hdz}{\hat{J}_{2z}}
\newcommand{\hup}{\hat{J}_1^+}
\newcommand{\hum}{\hat{J}_1^-}
\newcommand{\hdp}{\hat{J}_2^+}
\newcommand{\hdm}{\hat{J}_2^-}

\huge

\begin{center}
Semi-empirical cross-section formulas for electron-impact ionization and rate coefficient calculations
\end{center}

\vspace{0.5cm}

\large

\begin{center}
Djamel Benredjem$^{a,}$\footnote{djamel.benredjem@universite-paris-saclay.fr} and Jean-Christophe Pain$^{b,c}$
\end{center}

\normalsize

\begin{center}
\it $^a$Laboratoire Aim\'e Cotton, Universit\'e Paris-Saclay, Orsay, France\\
\it $^b$CEA, DAM, DIF, F-91297 Arpajon, France\\
\it $^c$Universit\'e Paris-Saclay, CEA, Laboratoire Mati\`ere en Conditions Extr\^emes,\\
\it 91680 Bruy\`eres-le-Ch\^atel, France\\
\end{center}

\vspace{0.5cm}

\begin{abstract}
We propose a semi-empirical formula for the cross section of ionization by electron impact. The formula involves adjustable parameters which are determined by comparison with measured or numerically calculated cross sections. In the latter case, the ions are perturbed by their environment which is a high-density plasma. As a consequence, the cross section is significantly modified. We investigate Be-like carbon, nitrogen and oxygen as well as aluminum ions. We also show that the formula is well-suited for interpolation and extrapolation. Knowing the cross section, we calculate the rate coefficient within the Boltzmann and Fermi-Dirac statistics. In the first case, the rate can be calculated analytically. In the second one, it can be expressed in terms of special functions, but the numerical evaluation is more convenient while providing accurate results. Our results are compared to experiment and to other calculations.
\end{abstract}

\section{Introduction}\label{Introduction}

Among the microscopic processes occurring in astrophysical and laboratory plasmas such as those encountered in inertial fusion studies, the ionization by electron impact plays an important role. Its knowledge is crucial when one is interested in radiative losses, spectral line shapes and opacity issues. In fact, the rate of such a process has an impact on ion-level populations. Moreover, the atomic structure as well as the absorption, emission and ionization, in media in non local thermodynamic equilibrium, must be treated carefully, in order to obtain accurate rates to be used in collisional-radiative calculations. Various methods are used to calculate the cross sections of several microscopic processes in many elements. For instance, Colgan \textit{et al.} studied the excitation and ionization of Si, Cl and Ar, in magnetic fusion and astrophysical modeling \cite{Colgan2008} while Pindzola \textit{et al.} \cite{Pindzola2008} considered the ionization of Ne and Au in hot and dense plasmas. In matter under extreme conditions, the atomic processes are often modified by plasma density effects, yielding in particular level shifts \cite{Belkhiri2015} or continuum lowering. In both cases, the cross section may be significantly modified \cite{Fontes1993}.

In dense and hot plasmas, we generally have several ion charges and the number of occupied excited states is important. Huge numbers of radiative transitions (up to billions in extreme cases) may then be involved in population kinetics. Statistical methods help reducing the number of collisional-radiative equations \cite{Ralchenko2016} but generally the resulting set remains large. As a consequence, calculations based on sophisticated methods (such as close coupling or R-matrix), where a cross section must be computed for each specific pair of initial-final states including channel mixing, cannot be used in intensive calculations. Therefore, semi-analytic methods are of great interest to obtain the rates of the microscopic processes required to solve such complex systems of equations. In a previous work on electron-impact excitation \cite{Pain2021}, we proposed analytical cross sections including plasma density effects within the screened hydrogenic model.

The literature about semi-empirical formulas for electron impact ionization (EII) cross sections is abundant (see for instance Refs. \cite{Gryzinski1965a,Drawin1977,Fujimoto1978,Casnati1982,Shevelko1983,Jakoby1987,Hombourger1993,Sobelman1995,Seah1998,Campos2007,Ralchenko2008,Bote2009,Jablonski2009}). For instance, the ones of Drawin \cite{Drawin1966}, Lotz \cite{Lotz1968} and Younger \cite{Younger1982} have been widely used in the past. More recently, Bernshtam \textit{et al.} \cite{Bernshtam2000} proposed for the direct ionization an empirical formula which is similar to the Lotz expression. It involves two parameters that depend on the orbital quantum number of the initial state. The two parameters are adjusted to fit experimental results. The general semi-empirical Bethe-Born formula has been used by Shevelko and Tawara to calculate the cross section of ionization from neutral to multiply-charged ions \cite{Shevelko1995}. The agreement with experiment is good for Mg, Fe and Cu but the theory underestimates the Xe and U cross sections. Also, a fitting formula for the rate coefficient was given by Voronov \cite{Voronov1997}. It involves five fitting parameters and provides values for ionization from the ground states of H to Ni$^{27+}$ and temperatures from 1 to 30000 eV. Llovet \textit{et al.} \cite{Llovet2014} (see also references therein) described the essentials of classical, semi-classical and quantum models, and made an extensive comparison of measured K, L and M shells of all elements from H to Es.

In this work, we propose a semi-empirical formula for the cross section. It involves four adjustable parameters that are determined by fitting cross sections measured in experiments on carbon, nitrogen and oxygen ions \cite{Fogle2008} and neutral aluminum \cite{Freund1990}. We investigate two expressions of the cross section. The first one is an extension of Kim's formula \cite{Kim1992}. The second one has the advantage of falling to zero when the incident electron energy is equal to the ionization energy. Both formulas easily fit experimental results and calculations using accurate atomic codes such HULLAC \cite{Barshalom2001} or FAC \cite{Gu2008}. Knowing the values of the adjustable parameters, we can (i) derive analytic expressions of the rate coefficients, (ii) calculate cross sections in energy ranges where no experimental results are available. Moreover, analytic formulas of the rate coefficient are very helpful in collisional-radiative calculations. 

In dense plasmas, the so-called continuum lowering may be important. This effect was thoroughly investigated experimentally \cite{Ciricosta2016} and theoretically \cite{Stewart-Pyatt1966,Ecker-Kroll1963,Zeng2020,Calisti2017,Benredjem2023}. In such media, the EII and the autoionization cross sections differ from that of isolated ions. In this work, we account for the continuum lowering in our analytic expression of the cross section.

Semi-empirical formulas for the EII cross section are presented in Section \ref{Cross sections}. The adjustable parameters are determined by a fit of experimental results in Be-like carbon, nitrogen and oxygen \cite{Fogle2008} and neutral aluminum \cite{Freund1990}. We also focus on an experiment performed at the Linac Coherent Light Source (LCLS) \cite{Ciricosta2016} on a high-density plasma. The plasma conditions are inferred from the experiment, and the density effects are taken into account in the calculation of the ionization energy. As a consequence, the cross sections vary by a substantial amount. In section \ref{Rate coefficient}, we calculate the rate coefficient using either Fermi-Dirac or Boltzmann statistics. While the rate associated with the extension of the Kim formula \cite{Kim1992} can be calculated easily, the one associated with our new cross section involves a Meijer G-function and the generalized hypergeometric function $_3F_3$ \cite{Prudnikov1990,Bateman1953}. However, the numerical calculation is accurate and fast.

\section{Cross sections}\label{Cross sections}

We first consider the following cross section:
\begin{equation}
    \sigma_1(E)=A\,\frac{\ln(E/E_i)}{E/E_i}+\sum_{p=1}^N\frac{B_p}{(E/E_i)^p},\label{Kim1}
\end{equation}
where $E$ is the incident-electron energy, $E_i$ the ionization energy and $\lbrace A, B_p\rbrace$ adjustable parameters. This formula is an extension of the Kim cross section \cite{Kim1992} in which $N=2$. The values of the parameters are determined by fitting accurate cross sections that are obtained with a reliable atomic code, \textit{e.g.} FAC or HULLAC, or determined experimentally. In all our calculations, the obtained cross section is satisfactory when $N=3$.

We also investigate the new expression:
\begin{equation}
    \sigma_2(E)=A\,\frac{\ln(E/E_i)}{E/E_i}\,\sum_{p=0}^N\frac{B_p}{(E/E_i)^p},\label{Kim2}
\end{equation}
with $B_0=1$ in all cases. Unlike $\sigma_1$, $\sigma_2$ fulfills the condition $\sigma_2(E_i)=0$, which is more satisfactory. In the following, we fit experimental or calculated cross sections to both analytic formulas, and show that in some cases $\sigma_2$ is more appropriate than $\sigma_1$. In most cases, the agreement between the two formulas is satisfactory. 

\subsection{Be-like C, N and O}\label{CNO}

\begin{figure}[!ht]
\begin{center}
\includegraphics[scale=.5]{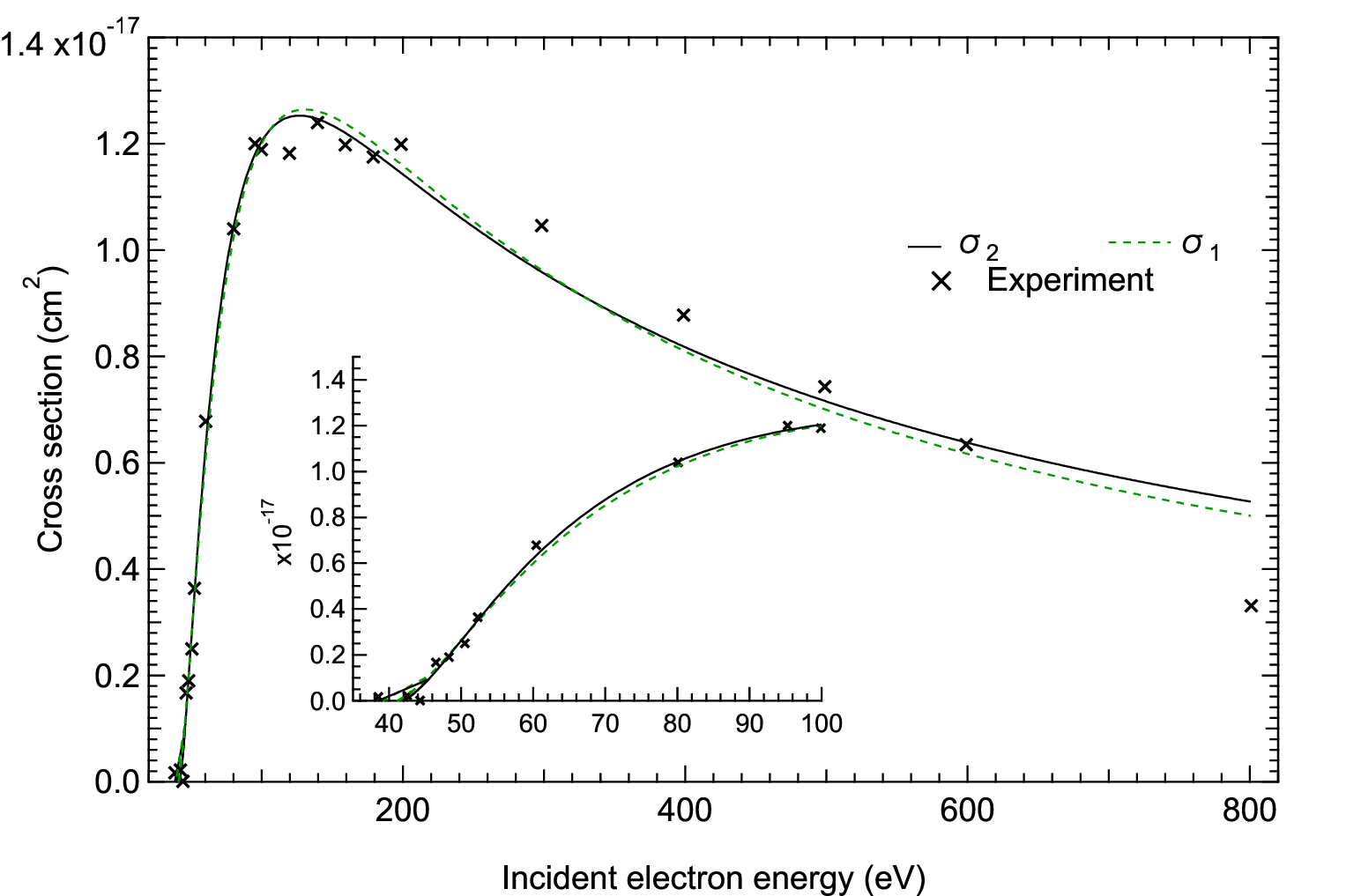}
\end{center}
\caption{Cross section of Be-like carbon (isolated ion). $\sigma_1$: Eq. (\ref{Kim1}), $\sigma_2$: Eq. (\ref{Kim2}), cross marks: experimental results (Ref. \cite{Fogle2008}). The measured ionization energy is $E_i=$38.5 eV. The insert gives an insight into the variation in the region where the cross section is increasing.}\label{BeC}
\end{figure}

\begin{figure}[!ht]
\begin{center}
\includegraphics[scale=.5]{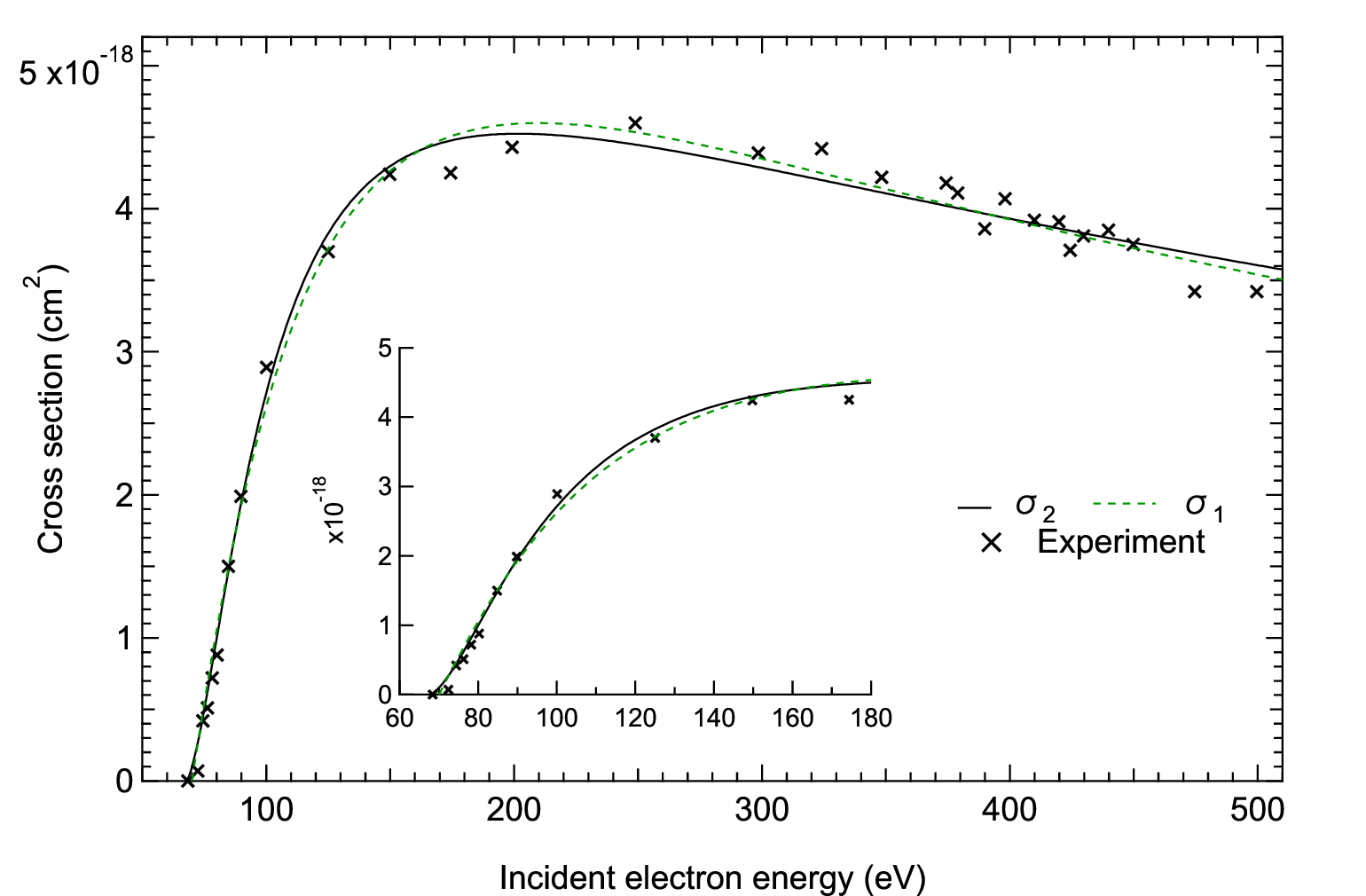}
\end{center}
\caption{Cross section of Be-like nitrogen (isolated ion). $\sigma_1$: Eq. (\ref{Kim1}), $\sigma_2$: Eq. (\ref{Kim2}), cross marks: experimental results (Ref. \cite{Fogle2008}). The measured ionization energy is $E_i=$68.4 eV. The insert gives an insight into the variation in the region where the cross section is increasing.}\label{BeN}
\end{figure}

\begin{figure}[!ht]
\begin{center}
\includegraphics[scale=.5]{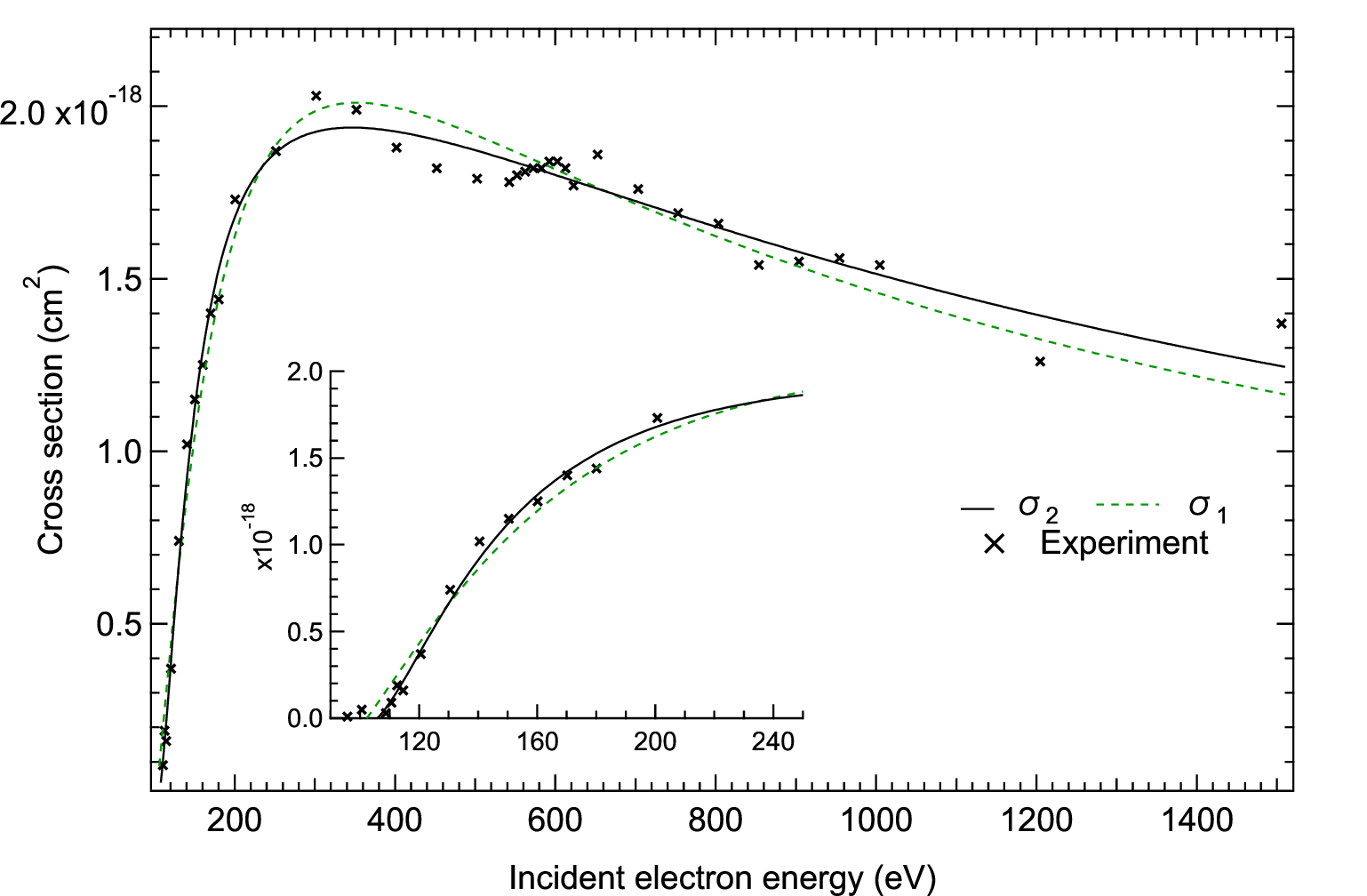}
\end{center}
\caption{Cross section of Be-like oxygen (isolated ion). $\sigma_1$: Eq. (\ref{Kim1}), $\sigma_2$: Eq. (\ref{Kim2}), cross marks: experimental results (Ref. \cite{Fogle2008}) The measured ionization energy is $E_i=$95.7 eV. The insert gives an insight into the variation in the region where the cross section is increasing.}\label{BeO}
\end{figure}

In this section, we concentrate on Be-like C, N and O ions. The cross sections obtained in experiments using the crossed-beams apparatus at Oak Ridge National Laboratory \cite{Fogle2008} have been fitted to both analytic expressions (Eq. (\ref{Kim1}) and Eq. (\ref{Kim2})), in order to obtain the values of the adjustable parameters. These values are given in Tables \ref{ParamCNO1} and \ref{ParamCNO2}. Figure \ref{BeC} shows the variation of the Be-like carbon cross section. We can see that $\sigma_1$ and $\sigma_2$ agree with each other and with experiment. The fit is also satisfactory in Be-like nitrogen (see Figure \ref{BeN}). In the case of Be-like oxygen (Figure \ref{BeO}), $\sigma_2$ fits very well the measured cross section while the agreement between $\sigma_1$ and the experiment is rather poor. In each figure we insert a subwindow in order to highlight the variation in the region where the cross section increases as well as the behavior near the threshold.

\begin{table}[!ht]
\begin{center}
\begin{tabular}{|l|r|r|r|}
\hline
& C$^{2+}\hspace{1.1cm}$&N$^{3+}\hspace{1.1cm}$& O$^{4+}\hspace{1.1cm}$ \\
 \hline
$A$ & $2.1523\times 10^{-17}$ &$1.1352\times 10^{-17}$& $7.4512\times 10^{-18}$ \\
$B_1$ & $4.3969\times 10^{-17}$ & $5.2241\times 10^{-18}$ & $-2.1181\times 10^{-18}$ \\
$B_2$ & $-1.1320\times 10^{-16}$ & $-1.4849\times 10^{-17}$ & $-1.7091\times 10^{-18}$ \\
$B_3$ & $6.9125\times 10^{-17}$ & $9.3973\times 10^{-18}$ & $3.6885\times 10^{-18}$ \\
\hline
\end{tabular}
\end{center}
\caption{\label{ParamCNO1} Values of the $\sigma_1$ adjustable parameters, in cm$^2$.}
\end{table}

\begin{table}[!ht]
\begin{center}
\begin{tabular}{|l|r|r|r|}
\hline
& C$^{2+}\hspace{.9cm}$ & N$^{3+}\hspace{.9cm}$ & O$^{4+}\hspace{.9cm}$ \\
 \hline
$A$ & $3.5737\times 10^{-17}$ &$1.4811\times 10^{-17}$& $8.0109\times 10^{-18}$\\
$B_1$ & $0.2659$ & $-1.0323$ & $-2.0252$ \\
$B_2$ & $-1.0816$ & $2.3133$ & $4.0820$ \\
$B_3$ & $-0.4359$ & $-2.1292$ & $-3.3828$ \\
\hline
\end{tabular}
\end{center}
\caption{\label{ParamCNO2} Values of the $\sigma_2$ adjustable parameters. $A$ is in cm$^2$ and the $B_p$s are dimensionless. $B_0$ is set equal to 1.}
\end{table}

\subsection{Neutral aluminum}\label{NeutralAl}
\begin{figure}[!ht]
\begin{center}
\includegraphics[scale=.5]{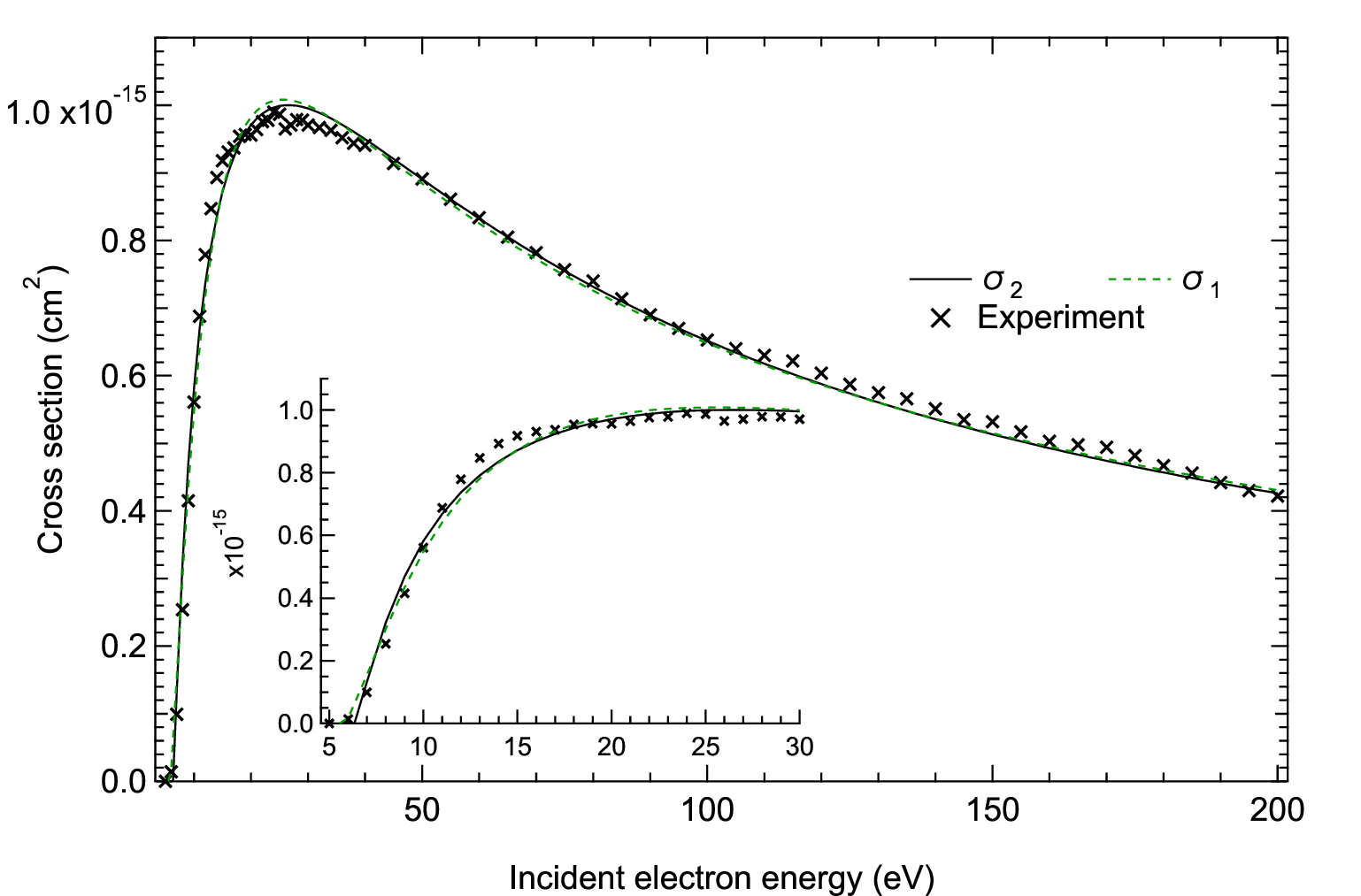}
\end{center}
\caption{Cross section of isolated Al I. $\sigma_1$: Eq. (\ref{Kim1}), $\sigma_2$: Eq. (\ref{Kim2}), cross marks: experimental results (Ref. \cite{Freund1990}). The measured ionization energy is $E_i=5$ eV. The insert gives an insight into the variation in the region where the cross section is increasing.}\label{Alneutral}
\end{figure}

Now we concentrate on neutral aluminum and rely on the measurements of Freund \textit{et al.} \cite{Freund1990}. The experiment used ion beams which are generated from a hot tungsten filament dc discharge, using a Colutron ion source (an ion gun system consisting of an ion source, an acceleration and focus lens system, a velocity filter and a beam decelerator) oriented with its horizontal axis. The accuracy of the measurements for single ionization is good (standard deviation below 6\% and equal to 5\% for aluminum). As above, the adjustable parameters are determined by a fitting procedure using the two analytic cross sections. The energy interval is limited to 200 eV. The values of the parameters are given in Table \ref{ParamAlneutral}. The cross sections are represented in Fig. \ref{Alneutral}. It is clear that $\sigma_1$ as well as $\sigma_2$ agree very well with each other and with the experimental results. The insert shows that close to the ionization energy, $\sigma_2$ is more satisfactory than $\sigma_1$. This is also the case in oxygen, as can be seen in Fig. \ref{BeO}.

\begin{table}[!ht]
\begin{center}
\begin{tabular}{|l|r|r|r|}
\hline
& $\sigma_1\hspace{.5cm}$ & $\sigma_2\hspace{.5cm}$ \\
 \hline
$A$ & $63.828$ &$49.116$ \\
$B_1$ & $-64.666$ & $-2.461$ \\
$B_2$ & $61.059$ & $3.651$ \\
$B_3$ & $3.399$ & $-2.724$ \\
\hline
\end{tabular}
\end{center}
\caption{\label{ParamAlneutral}Values of the adjustable parameters for neutral aluminum. $\sigma_1$: all the parameters are in units of 10$^{-16}$ cm$^2$, $\sigma_2$: $A$ is in units of 10$^{-16}$ cm$^2$ while the $B_p$s are dimensionless.}
\end{table}

\subsection{Aluminum ions}\label{Aluminum}
\begin{figure}[!ht]
\begin{center}
\includegraphics[scale=.5]{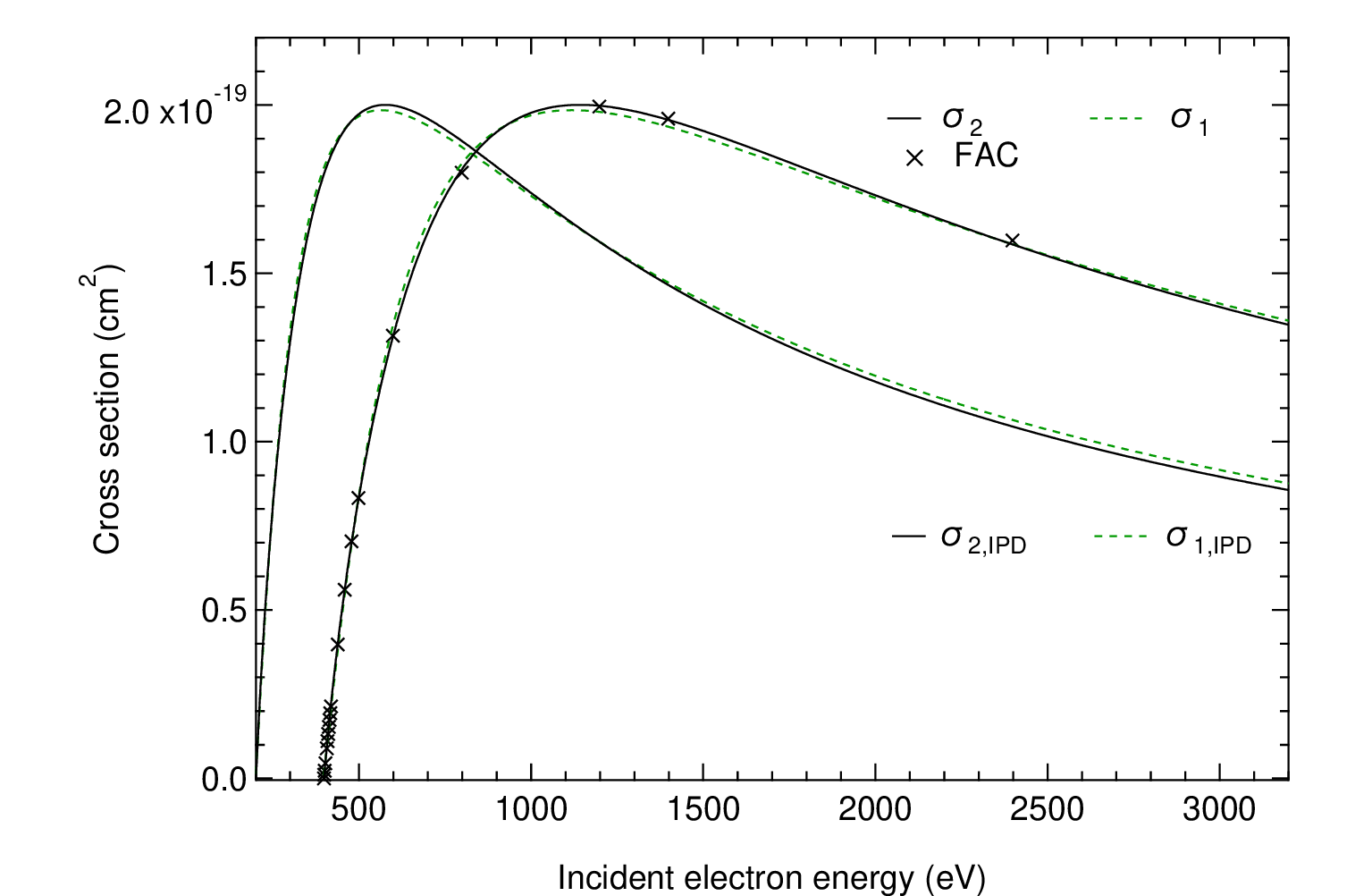}
\end{center}
\caption{Cross section of Be-like aluminum. FAC (FAC code calculations), $\sigma_1$ (Eq. (\ref{Kim1})) and $\sigma_2$ (Eq. (\ref{Kim2})) are for an isolated ion. $\sigma_{1,\rm IPD}$ and $\sigma_{2,\rm IPD}$ are the corresponding cross sections for a surrounded ion.}\label{BeAl}
\end{figure}

\begin{figure}[!ht]
\begin{center}
\includegraphics[scale=.5]{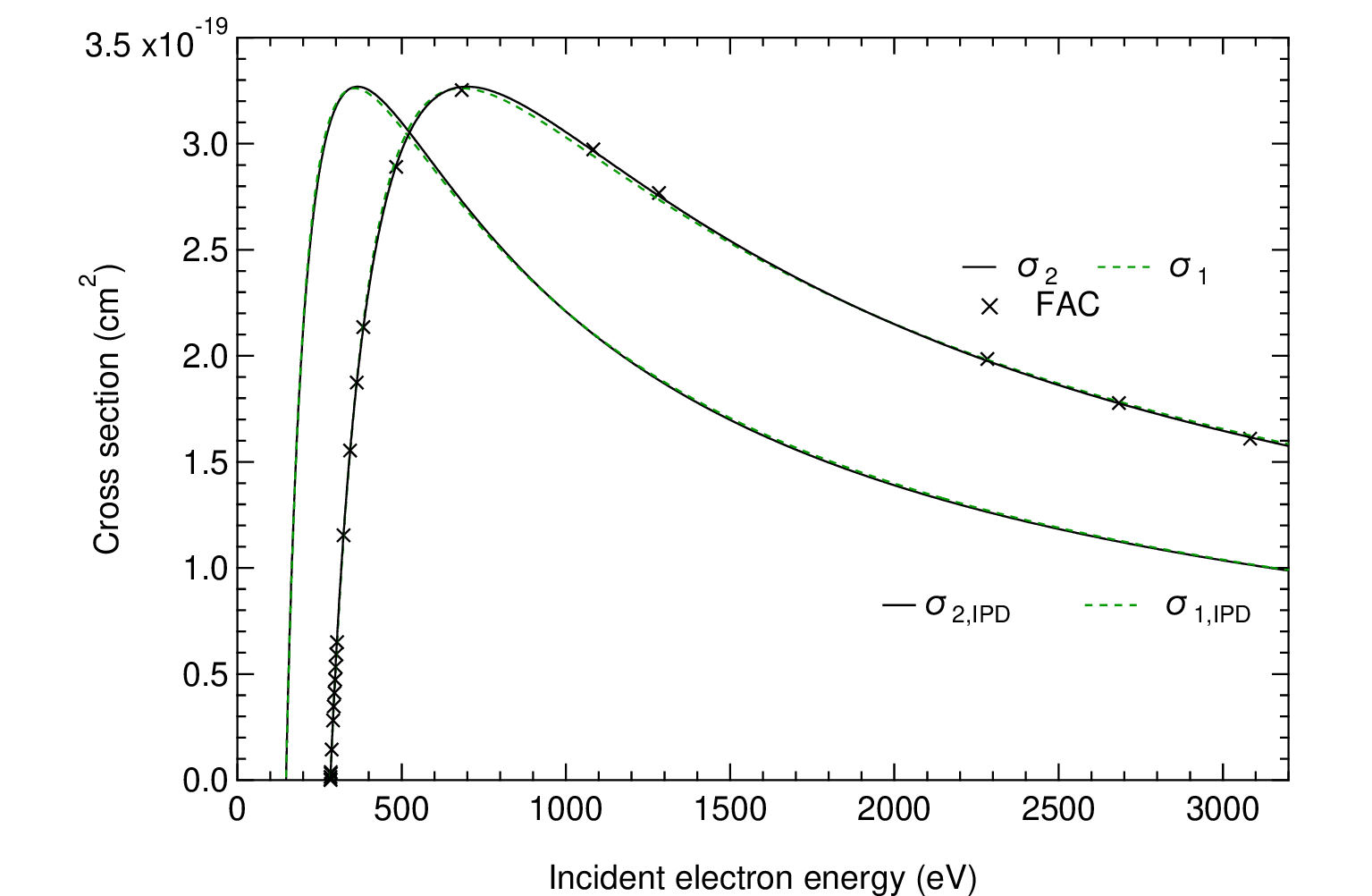}
\end{center}
\caption{Cross section of C-like aluminum. FAC (FAC code calculations), $\sigma_1$ (Eq. (\ref{Kim1})) and $\sigma_2$ (Eq. (\ref{Kim2})) are for an isolated ion. $\sigma_{1,\rm IPD}$ and $\sigma_{2,\rm IPD}$ are the corresponding cross sections for a surrounded ion.}\label{CAl}
\end{figure}

\begin{figure}[!ht]
\begin{center}
\includegraphics[scale=.5]{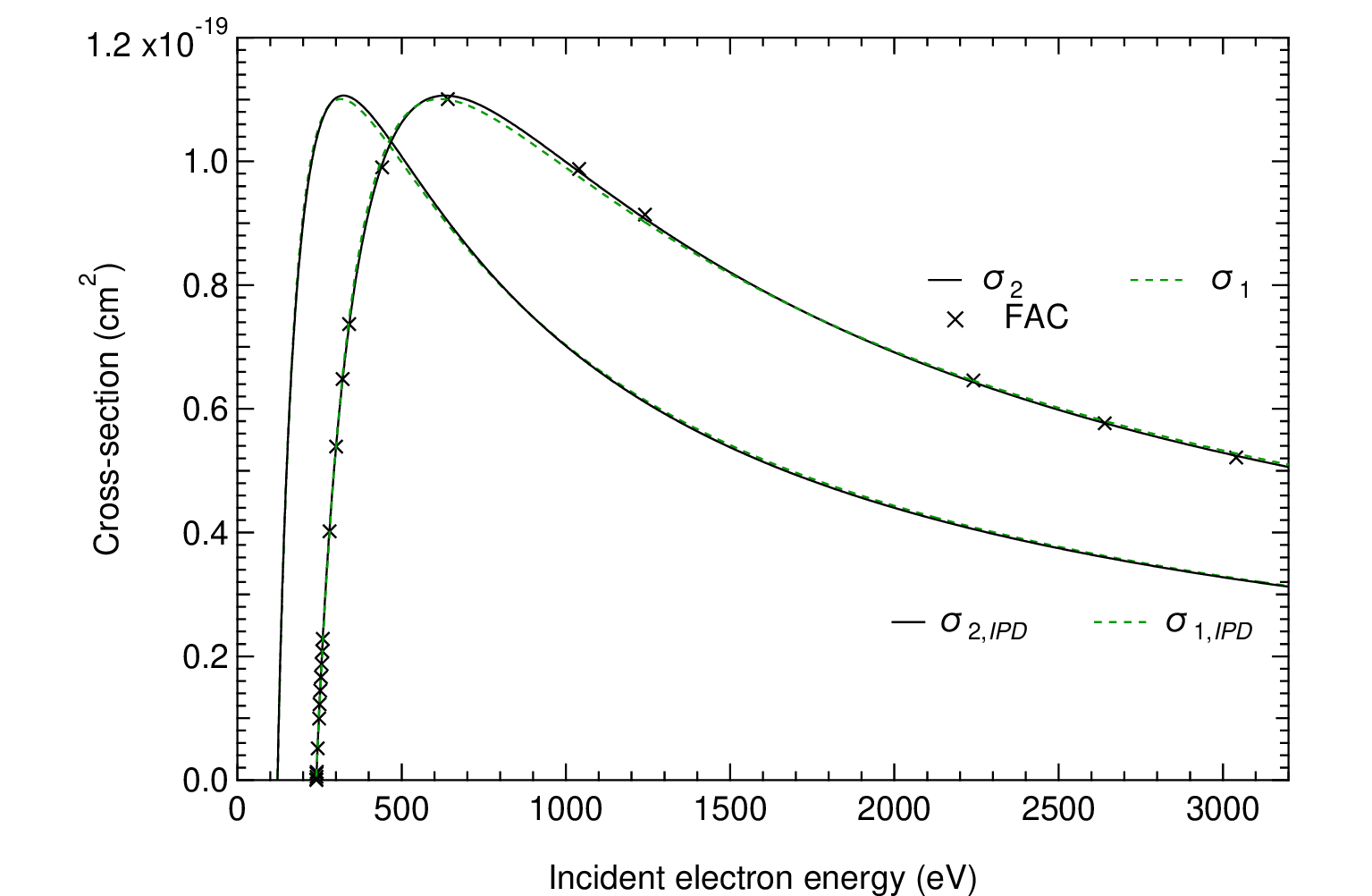}
\end{center}
\caption{Cross section of N-like aluminum. FAC (FAC code calculations), $\sigma_1$ (Eq. (\ref{Kim1})) and $\sigma_2$ (Eq. (\ref{Kim2})) are for an isolated ion. $\sigma_{1,\rm IPD}$ and $\sigma_{2,\rm IPD}$ are the corresponding cross sections for a surrounded ion.}\label{NAl}
\end{figure}

Due to a lack of measured cross sections in aluminum ions, we choose to fit FAC cross sections for isolated ions by $\sigma_1$ and $\sigma_2$. In this section, we restrict ourselves to transitions between the ground levels of the two involved ions. The obtained parameter values are given in Table \ref{AdjustParam}. Figures \ref{BeAl}$-$\ref{NAl} show the cross sections of Be-, C- and N-like ions, respectively. We can see that the fit is very satisfactory in the whole energy range. 

In a dense hot plasma, the ionization potential is lowered due to screening by charged particles. This effect, known as the continuum lowering or ionization potential depression (IPD), can affect the cross section. In this work, we account for this effect on the EII cross section of aluminum ions. We consider a plasma under the conditions of the LCLS experiment \cite{Ciricosta2016}, \textit{i.e.}, a mass density and an electron temperature equal to 2.7 g/cm$^3$ and 50 eV, respectively. In the case of the Be-like ion, the measured IPD is 201 eV yielding an ionization energy of 197 eV. At such a density, the effect on the new cross sections ($\sigma_{\rm 1,IPD}$ and $\sigma_{\rm 2,IPD}$) is expected to be important. In fact, we can see that the difference with the isolated-ion cross sections is important. In addition, $\sigma_{\rm 1,IPD}$ and $\sigma_{\rm 2,IPD}$, are very close to each other.

\begin{table}
\begin{center}
\begin{tabular}[h!]{cc}
\begin{tabular}[t]{|l|r|r|r|}
\hline
& Al$^{9+}\hspace{.2cm}$ & Al$^{7+}\hspace{.2cm}$ & Al$^{6+}\hspace{.2cm}$\\
 \hline
$A$ & 3.8036 & 4.8011 & 1.8697 \\
$B_1$ & 3.8874 & 7.0539 & 2.1994 \\
$B_2$ & -7.6343 & -9.0867& -3.3799 \\
$B_3$ & 3.7610 & 2.0461 & 1.1862 \\
\hline
\end{tabular} &
\hspace{1cm}
\begin{tabular}[t]{|l|r|r|r|}
\hline
 & Al$^{9+}\hspace{.2cm}$& Al$^{7+}\hspace{.2cm}$ & Al$^{6+}\hspace{.2cm}$\\
\hline
$A$ & 4.5440 & 6.4397 & 2.3624 \\
$B_1$ & 1.5595 & 1.8190 & 1.5729 \\
$B_2$ & -3.5505 & -2.7980 & -2.8827 \\
$B_3$ & 2.0352& 1.6410 & 1.6803 \\
\hline
\end{tabular} \tabularnewline
\end{tabular}
\end{center}
\caption{\label{AdjustParam}Values of the parameters. $A$ is in units of $10^{-19}$ cm$^2$, left ($\sigma_1$): the $B_l$s are in units of $10^{-19}$ cm$^2$, right ($\sigma_2$): the $B_l$s are dimensionless.}
\end{table}

Our formulas can be used to obtain accurate cross sections in energy ranges where experimental data are not available. They can also provide rate coefficients at various temperatures and densities, by using analytical or numerical calculations (see Section \ref{Rate coefficient}).

\subsection{$\sigma_2$ vs $\sigma_1$}
In most cases, $\sigma_1$ and $\sigma_2$ agree very well with each other if their respective parameters are obtained by a fit of the same set of measured or calculated cross sections. However, as $\sigma_2$ is the product of a logarithm and a polynomial, the ionization threshold occurs at the ionization energy. Moreover, as we will see below, $\sigma_2$ is more appropriate for extrapolation or interpolation than $\sigma_1$. 

\begin{figure}[!ht]
\begin{center}
\includegraphics[scale=.5]{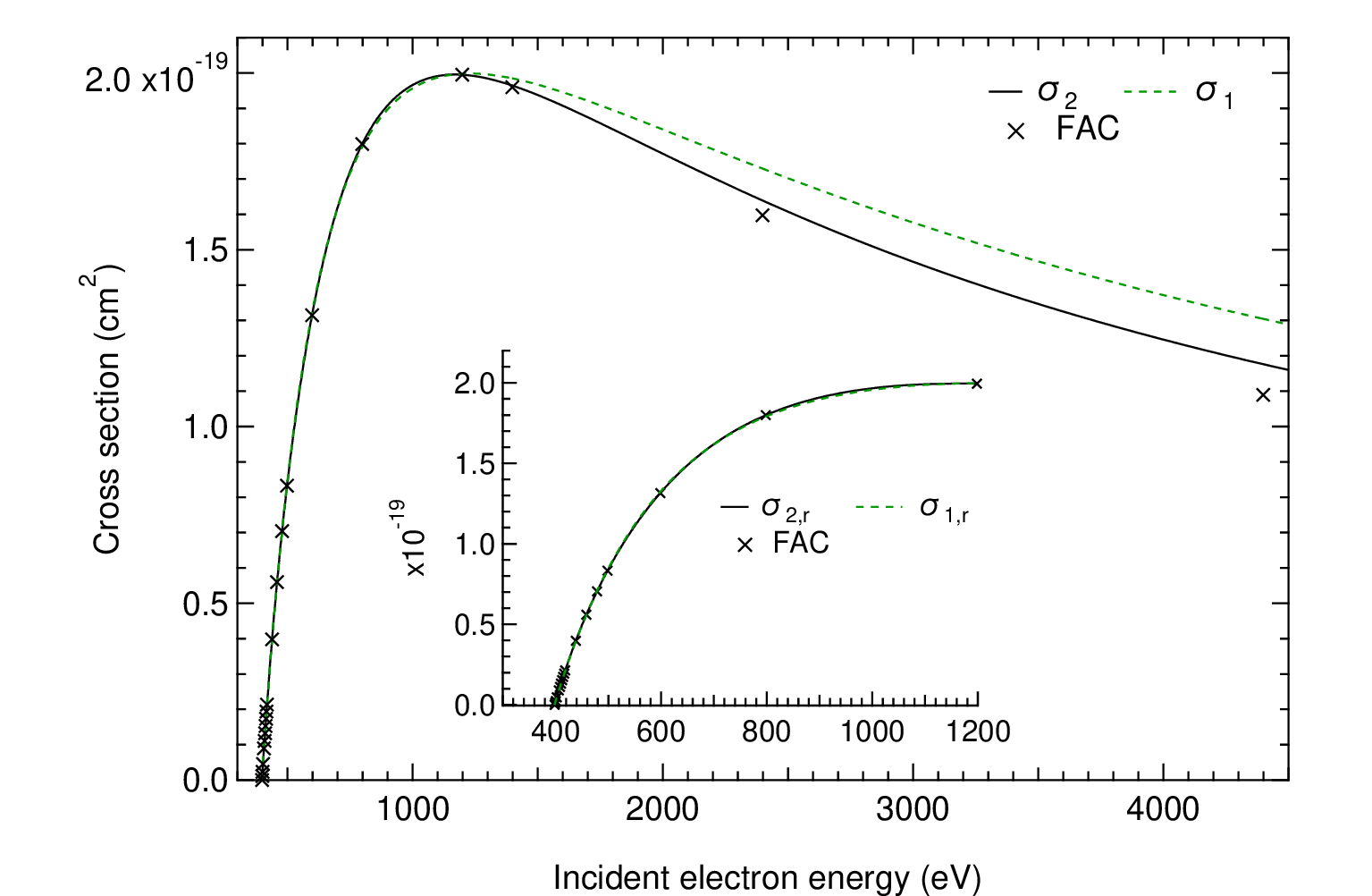}
\end{center}
\caption{Cross sections of an isolated Be-like aluminum ion vs incident electron energy. FAC (FAC code calculations), $\sigma_1$ (Eq. (\ref{Kim1})) and $\sigma_2$ (Eq. (\ref{Kim2})). The index $r$ in the insert refers to the restricted energy interval [398,1200] eV involved in the fitting procedure.}\label{BeAl2}
\end{figure}

\begin{figure}[!ht]
\begin{center}
\includegraphics[scale=.5]{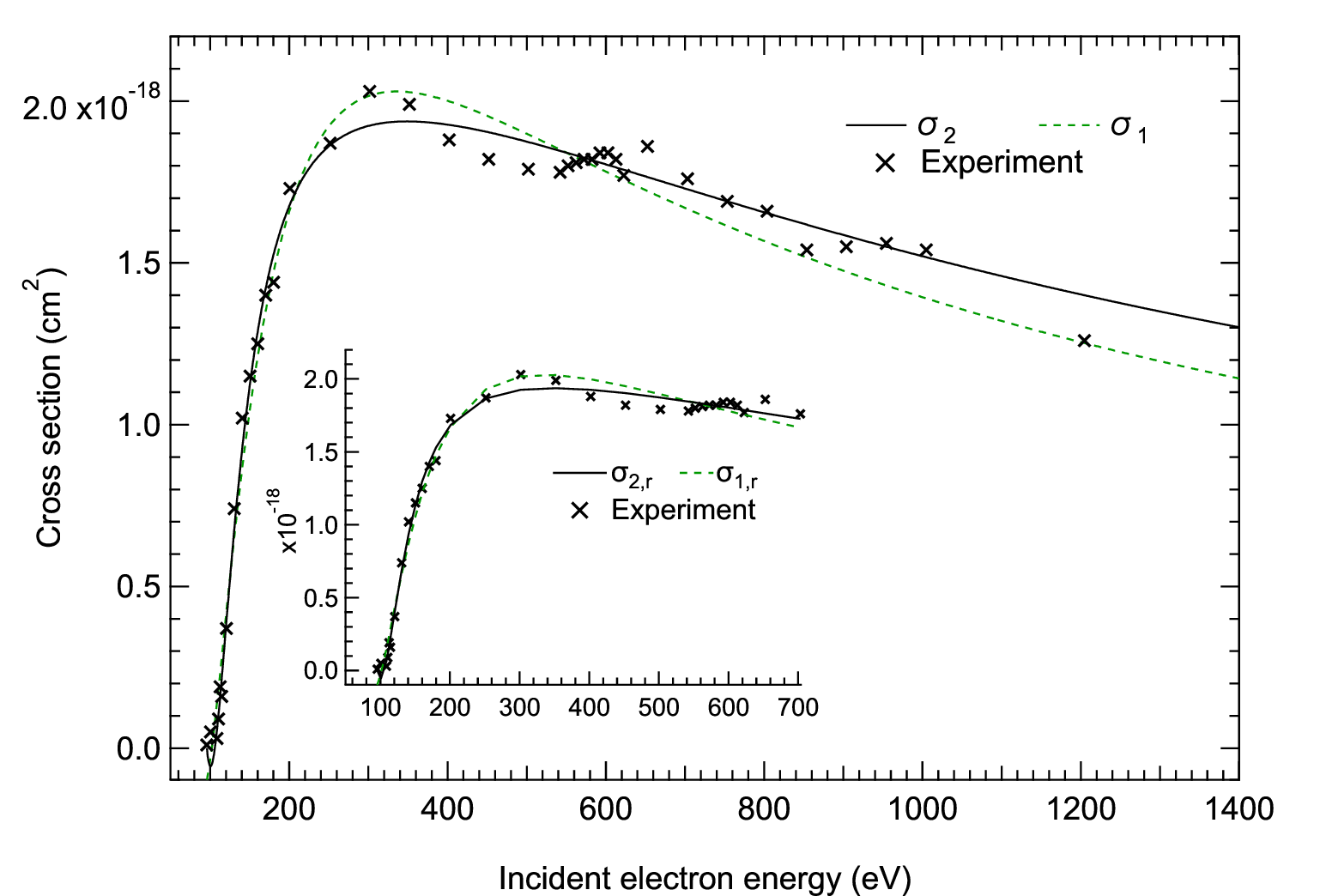}
\end{center}
\caption{Cross sections of an isolated Be-like oxygen ion vs incident electron energy. Experiment: Ref. \cite{Fogle2008}, $\sigma_1$ (Eq. (\ref{Kim1})) and $\sigma_2$ (Eq. (\ref{Kim2})). The index $r$ in the insert refers to the restricted energy interval [95-700] eV involved in the fitting procedure.}\label{BeO2}
\end{figure}

Let us first consider Be-like aluminum. As shown in Fig. \ref{BeAl}, the variation of the FAC cross section in the [400-3200] eV energy interval is well reproduced by $\sigma_1$ and $\sigma_2$. If the adjustable parameters are now obtained within the smaller interval [400-1200] eV (see Fig. \ref{BeAl2}, the insert), $\sigma_1$ and $\sigma_2$ still agree very well with FAC results. Both cross sections are then extrapolated to the rest of the interval, \textit{i.e.}, [1200-4000] eV while using the same respective parameters. As a result, $\sigma_2$ is still in good agreement with FAC cross sections while $\sigma_1$ shows a substantial discrepancy. In this case, it is clear from Fig. \ref{BeAl2} that $\sigma_2$ is better suited for extrapolation than $\sigma_1$. Figure \ref{BeO} showed that, for Be-like oxygen, both $\sigma_2$ and $\sigma_1$ agree fairly well with experimental results. If the analytic fit to the experimental cross sections is realized within the smaller energy interval [95-700] eV (see Fig. \ref{BeO2}), $\sigma_1$ and $\sigma_2$ present the same level of agreement with the measured cross sections. But if we extend $\sigma_2$ and $\sigma_1$ to the whole energy interval while using the same values for the respective parameters, we can see that $\sigma_2$ is still in good agreement with experiment while $\sigma_1$ shows a substantial discrepancy. Through these two comparisons, we can conclude that the new cross section, $\sigma_2$, is more satisfactory than $\sigma_1$. 

In the last section, we will derive analytic rate coefficients for both cross sections, as long as one deals with a Boltzmann distribution of free electrons. We also present expressions of the rate coefficient when the Fermi-Dirac distribution is more appropriate than the Boltzmann one.

\section{Rate coefficient}\label{Rate coefficient}
\subsection{Distribution of the free electrons}

In Fermi-Dirac statistics, the distribution of free electrons is given by
\begin{equation}
    \rho(E)=\frac{1}{D}\frac{\sqrt{E}}{e^{(E-\mu)/k_{\rm B}T_e}+1},\label{FD-distribution}
\end{equation}
where $T_e$ is the electron temperature, $k_{\rm B}$ the Boltzmann constant and $\mu$ the chemical potential. $D$ is the normalization constant given by
$$D=(k_{\rm B}T_e)^{3/2}\Gamma(3/2)F_{1/2}(\eta),$$
where $\Gamma$ is the usual Gamma function, $\eta=\mu/(k_{\rm B}T_e)$ and $F_{1/2}$ the complete Fermi-Dirac integral, defined by
\begin{equation*}
    F_{1/2}(\eta)=\frac{2}{\sqrt{\pi}}\int_0^{\infty}\frac{\sqrt{t}}{1+e^{t-\eta}}\,dt.
\end{equation*}
In hot plasmas, the factor $\displaystyle \sqrt{E}/(e^{(E-\mu)/k_{\rm B}T_e}+1)$ often reduces to $\sqrt{E}~e^{-(E-\mu)/k_{\rm B}T_E}$, which allows to use the Boltzmann distribution. As a consequence, the Fermi-Dirac integral becomes
\begin{equation*}
    F_{1/2}(\eta)\to \frac{2}{\sqrt{\pi}}\,e^{\eta}\int_0^{\infty}\sqrt{t}\,e^{-t}\,dt=e^{\eta},
\end{equation*}
and the Boltzmann distribution reads
\begin{equation}
    \rho'(E)=\frac{2}{\sqrt{\pi}}\frac{1}{(k_{\rm B}T_e)^{3/2}}\sqrt{E}e^{-E/(k_{\rm B}T_e)}.\label{B-distribution}
\end{equation}

\subsection{Calculation of the reduced chemical potential $\eta$}

A straightforward calculation of the number of free electrons in a given volume allows one to write the Fermi-Dirac integral as

\begin{equation}
    F_{1/2}(\eta)=\frac{(4\pi)^{3/2}}{2}\,\left(\frac{\mathrm{Ryd}}{k_{\rm B}T_e}\right)^{3/2} N_e,
\end{equation}
where $N_e$ is the electron density and $\mathrm{Ryd}$ the Rydberg constant. The equation can be solved numerically to obtain $\eta$. It is worth mentioning that Blakemore \cite{Blakemore1982} proposed accurate approximate solutions. For instance, in a gas at $k_{\rm B}T_e$=50 eV and $N_e=3.47\times 10^{23}$ cm$^{-3}$, the exact solution is -1.75927 while two solutions are given in Ref. \cite{Blakemore1982}. The first one: 
\begin{equation}
    \frac{\ln[F_{1/2}(\eta)]}{1-F_{1/2}^2(\eta)}+\frac{\left(3\sqrt{\pi}F_{1/2}(\eta)/4\right)^{2/3}}{1+\left[0.24+1.08\,\left(3\sqrt{\pi}F_{1/2}(\eta)/4\right)^{2/3}\right]^{-2}}
\end{equation}
gives $\eta=-1.76386$, and the second one:
\begin{equation}
    \ln[F_{1/2}(\eta)]+F_{1/2}(\eta)\,\left[64+0.05524\,F_{1/2}(\eta)\,\left(64+\sqrt{F_{1/2}(\eta)}\right)\right]^{-1/4} 
\end{equation}
provides the exact value $\eta=-1.75927$.

\subsection{Rate coefficient calculation}

The EII rate coefficient reads
\begin{equation}
    q=\int_{E_i}^{\infty} \sigma(E)\,v\,\rho(E)\,dE,
\end{equation}
where $v=\sqrt{2E/m_e}$ is the velocity of the incident electron, with $m_e$ the electron mass. 

If the electron gas obeys the Fermi-Dirac statistics, the rate coefficient reads
\begin{equation*}
    q_{\rm{FD}}=\frac{1}{D}\sqrt{\frac{2}{m_e}}\int_{E_i}^{\infty}\sigma(E)\,\sqrt{E}\frac{\sqrt{E}}{e^{(E-\mu)/k_{\rm B}T}+1}\,dE.
\end{equation*}
Having $\Gamma(3/2)=\sqrt{\pi}/2$ and defining the dimensionless variables $t=E/(k_{\rm B}T_e)$ and $b=E_i/(k_{\rm B}T_e)$, we easily obtain
\begin{equation}
    q_{\rm{FD}}=\sqrt{\frac{8\,k_{\rm B}T_e}{\pi\,m_e}}\frac{1}{F_{1/2}(\eta)}\int_{b}^{\infty}\sigma_i(t)\,t\frac{1}{e^{t-\eta}+1}\,dt.
\end{equation}
The above integral cannot be calculated analytically. However, its numerical value can be obtained with great accuracy. 

In the case where the Boltzmann statistics is relevant, the rate coefficient becomes
\begin{equation}
    q_{\rm{B}}=\sqrt{\frac{8\,k_{\rm B}T_e}{\pi\,m_e}}\,\int_b^{\infty}t\,\sigma(t)\, e^{-t}\,dt.
\end{equation}
This integral can be calculated analytically.

\subsection{$\sigma_1$ case}

Focusing on $\sigma_1$ (Eq. (\ref{Kim1})), the rate coefficient reads
\begin{eqnarray*}
    q_{1,\rm{B}}&=&\sqrt{\frac{8\,k_{\rm B}T_e}{{\pi\,m_e}}}\,\int_b^{\infty}t\,\sigma_1(t)\,e^{-t}\,dt\nonumber\\
    &=&\sqrt{\frac{8\,k_{\rm B}T_e}{{\pi\,m_e}}}\,b\,\int_b^{\infty}
    \left[A~\ln\left(\frac{t}{b}\right)+\sum_{p=1}^3\frac{B_p}{(t/b)^{p-1}}\right]\,e^{-t}\,dt\\\nonumber
    &=&\sqrt{\frac{8\,k_{\rm B}T_e}{{\pi\,m_e}}}\,b\,\int_1^{\infty}\left[A~\ln(x)+\sum_{p=1}^3\frac{B_p}{x^{p-1}}\right]\,e^{-b\,x}\,b\,dx,
\end{eqnarray*}
where $x=t/b$. The contribution involving the logarithm is integrated by parts, giving $A\,E_1(b)$, where $E_1$ is the exponential integral (see Ref. \cite{Abramowitz1972}). We then obtain
\begin{equation}
    q_{1,\rm{B}}=\sqrt{\frac{8\,k_{\rm B}T_e}{\pi\,m_e}}\,b
    \left\lbrace A\,E_1(b)+b\sum_{p=1}^3B_p\,E_{p-1}(b)\right\rbrace,
\end{equation}
where
\begin{equation*}
    E_q(b)=\int_1^{\infty}\frac{e^{-b\,x}}{x^q}\,dx.
\end{equation*}
We have $E_0(b)=e^{-b}/b$, $E_1(b)=\Gamma(0,b)$ and $E_2(b)=e^{-b}-b\,\Gamma(0,b)$, where $\Gamma(\cdot,b)$ is an incomplete Gamma function \cite{Abramowitz1972}. The rate can also be written in terms of $\Gamma(0,b)$ only:
\begin{equation}
    q_{1,\rm{B}}=\sqrt{\frac{8\,k_{\rm B}T_e}{\pi\,m_e}}\,b
    \left[ e^{-b}\left (B_1+bB_3\right )+\Gamma(0,b)\left(A+bB_2-b^2B_3\right )\right].
\end{equation}

\subsection{$\sigma_2$ case}

Considering $\sigma_2$ (see Eq. (\ref{Kim2})), the rate coefficient reads
\begin{eqnarray*}
    q_{2,\rm{B}}&=&\sqrt{\frac{8\,k_{\rm B}T_e}{{\pi\,m_e}}}\,\int_b^{\infty}t\,\sigma_2(t)\,dt\nonumber\\
    &=&\sqrt{\frac{8\,k_{\rm B}T_e}{{\pi~m_e}}}\,b~A\sum_{p=0}^3B_p~\int_b^{\infty}\ln\left(\frac{t}{b}\right)\frac{1}{(t/b)^{p}}\,e^{-t}\,dt\nonumber\\
    &=&\sqrt{\frac{8\,k_{\rm B}T_e}{{\pi~m_e}}}\,b^2\,A\sum_{p=0}^3B_p~\int_1^{\infty}~\ln(x)\frac{1}{x^p}\,e^{-b\,x}\,dx.
\end{eqnarray*}
In this case, the rate coefficient can be expressed in terms of the generalized integro-exponential functions \cite{Milgram1985,MacLeod2002,Luke1969}:
\begin{equation*}
    E_p^j(b)=\frac{1}{\Gamma(j+1)}\int_1^{\infty} [\ln(x)]^j\,\frac{1}{x^p}e^{-b\,x}\,dx.
\end{equation*}
We then have
\begin{equation}
    q_{2,\rm{B}}=\sqrt{\frac{8\,k_{\rm B}T_e}{\pi\,m_e}}\,b^2\,A\sum_{p=0}^3B_p\,\Gamma(p+1)E_p^1(b),
\end{equation}
where the functions $E_p^1$ are easily calculated. We have in particular
\begin{eqnarray*}
    E_0^1(b)&=&\int_1^{\infty}\ln(x)\,e^{-bx}\,dx=\frac{1}{b}\,\Gamma(0,b),
\end{eqnarray*}
and
\begin{equation}
    E_1^1(b)=\int_1^{\infty}\frac{\ln(x)}{x}\,e^{-bx}\,dx=G_{2,3}^{3,0}\left(b\left|\begin{array}{c}
    1,1\\
    0, 0,0
    \end{array}\right.\right),
\end{equation}
where $G$ is a Meijer $G$-function \cite{Prudnikov1990}.
The other integrals involved in the rate coefficients are easily obtained by using the recurrence relation \cite{Milgram1985}:
\begin{equation}
    E_p^1(b)=\frac{b\,E_{p-1}^1(b)-E_{p}^0(b)}{1-p}=\frac{b\,E_{p-1}^1(b)-E_{p}(b)}{1-p}\hspace{2cm} p\neq 1.
\end{equation}

\subsection{Rate coefficient versus temperature}
\begin{figure}[!ht]
\begin{center}
\includegraphics[scale=.5]{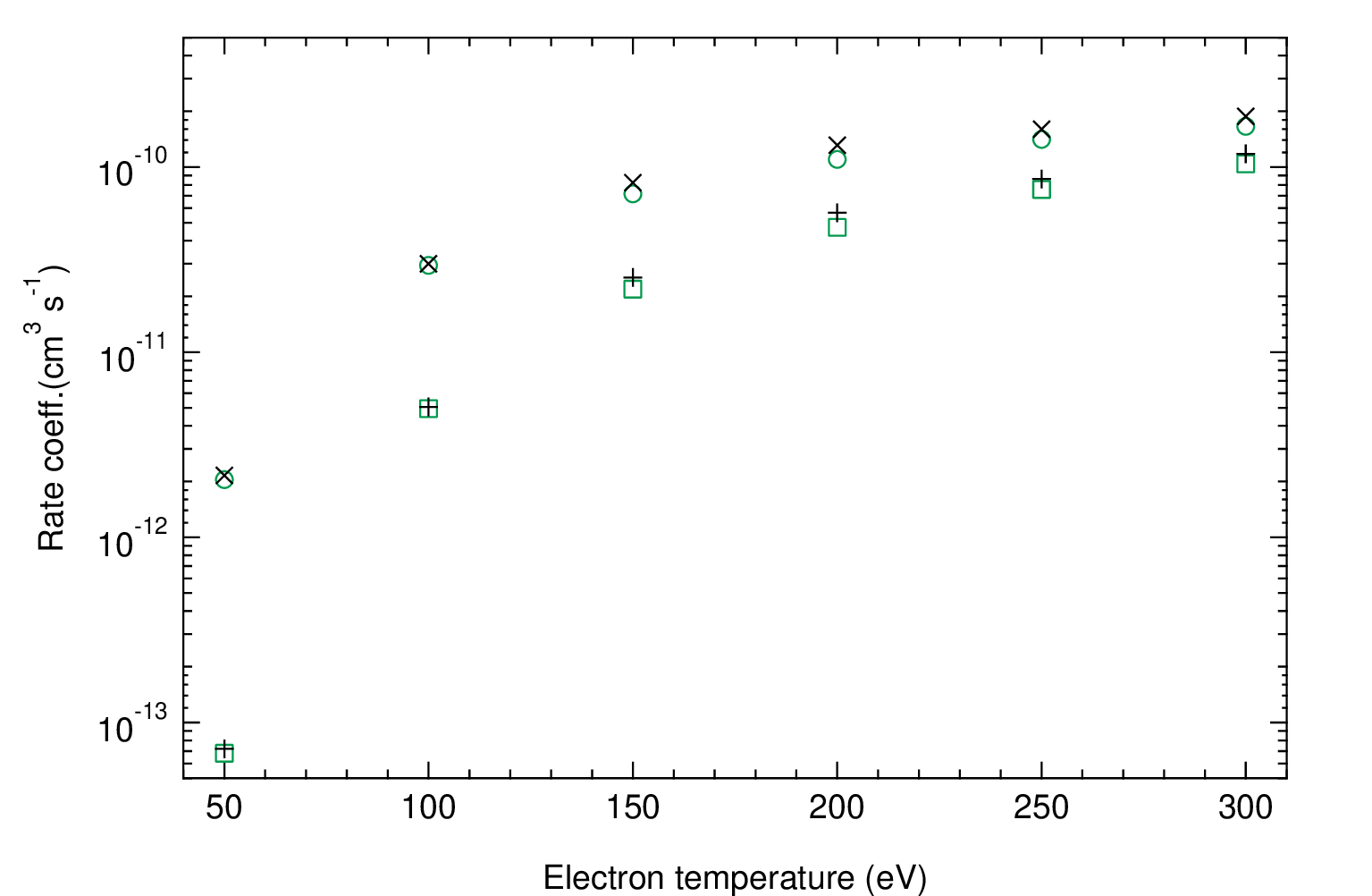}
\caption{Rate coefficient of Be-like aluminum vs electron temperature. Boltzmann statistics: isolated ($\textcolor{green}{\square}$), surrounded ion ($\textcolor{green}{\bigcirc}$). Fermi-Dirac statistics: isolated ($+$), surrounded ion ($\times$). Mass density: 2.7 g/cm$^3$.\label{Rate_Coef_BeAl}}
\end{center}
\end{figure}

\begin{figure}[!ht]
\begin{center}
\includegraphics[scale=.5]{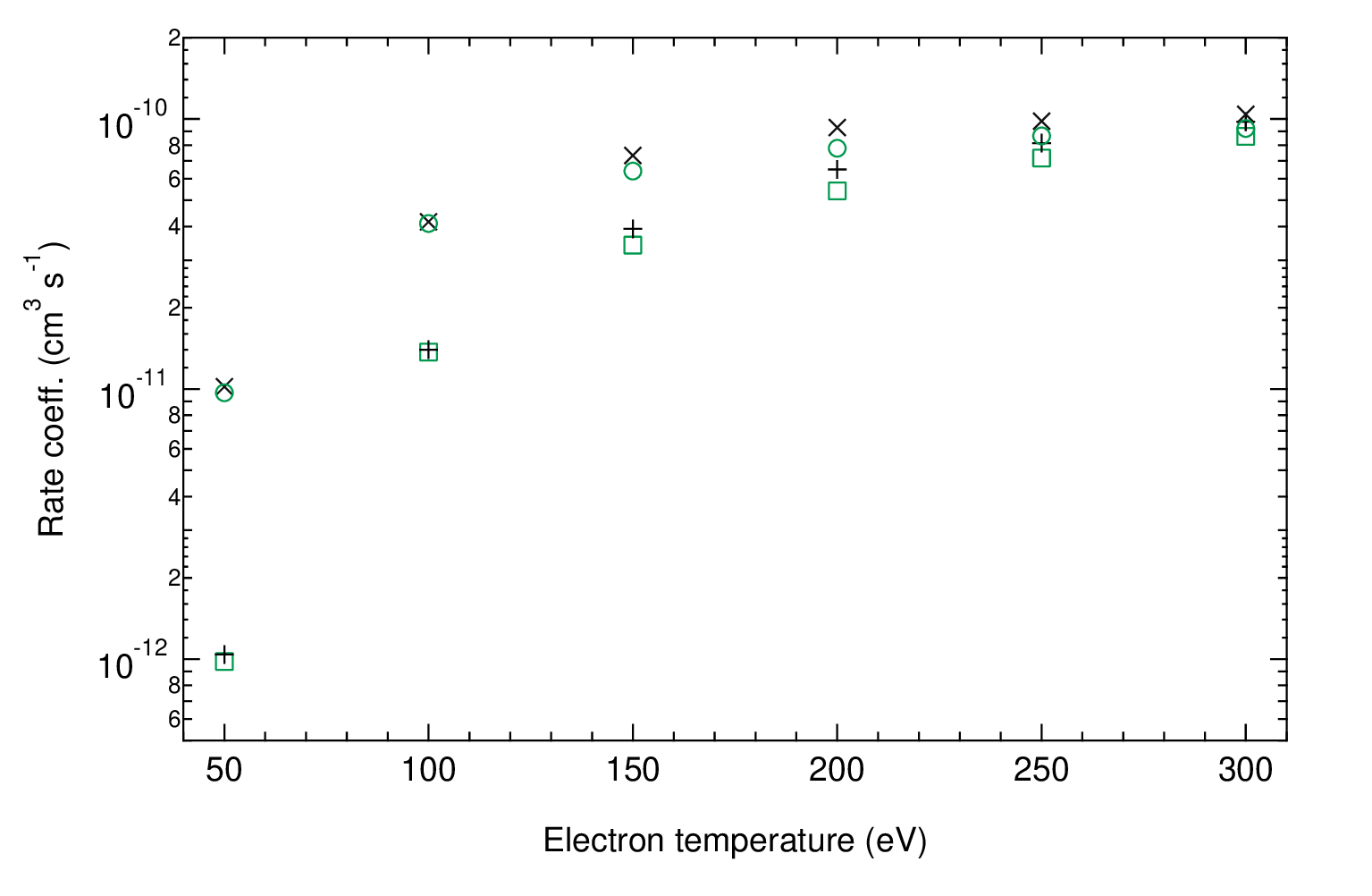}
\caption{Rate coefficient of N-like aluminum vs electron temperature. Boltzmann statistics: isolated ($\textcolor{green}{\square}$), surrounded ion ($\textcolor{green}{\bigcirc}$). Fermi-Dirac statistics: isolated ($+$), surrounded ion ($\times$). Mass density: 2.7 g/cm$^3$.\label{Rate_Coef_NAl}}
\end{center}
\end{figure}

In this section, we focus on the rate coefficients of aluminum ions. We first restrict ourselves to the ionization between the ground states involved in the transition. The density and temperature, inferred from the experiment at LCLS \cite{Ciricosta2016}, are respectively 2.7 g/cm$^3$ and 50 eV. In that case, the IPD is important and must therefore be taken into account in the rate calculation. In practice, the lower bound of the integral giving the rate is $b=E_i/(k_{\rm B}T_e)$, where the ionization energy $E_i$ must include the IPD. Due to the lack of measured IPDs at other temperatures, we use the formula of Ecker-Kr\"oll \cite{Ecker-Kroll1963}. Figure \ref{Rate_Coef_BeAl} shows the variation of the Be-like aluminum rate coefficient for both isolated and surrounded (perturbed) ions. We can see that the continuum lowering plays an important role. The number of free electrons increases with the temperature. As a consequence, the electron density increases and the difference between the two statistical distributions becomes significant. The N-like ion (Figure \ref{Rate_Coef_NAl}) behaves similarly to the Be-like one.

\subsection{Comparisons}
We calculated the EII rate of isolated aluminum ions at the temperatures investigated by Voronov \cite{Voronov1997} (see also Ref. \cite{Lennon1988}). These rates are given in Table \ref{Comparison1}, where we have three entries for a given ion and temperature. The first one accounts for ionization from the ground state to a small set of final states comprising the ground state and a few excited states of the next ion stage. In this case, the fit of $\sigma_2$ to FAC cross sections yields the adjustable parameters in Table \ref{AdjustParam3}. The second entry is the rate of ionization between the two ground states (adjustable parameters are in Table \ref{AdjustParam}). The third one is the value tabulated in Ref. \cite{Voronov1997}. We can see that 

\begin{itemize}
    \item considering only the ground-to-ground channel yields a poor agreement with Voronov results,
    \item additional channels involving excited states of the final ion improve the agreement. For example, in N-like aluminum, if we add the three channels: $\displaystyle 1s^22s^22p^3~(^4S_{3/2}) \to 1s^22s^22p^2~(^3P_{1,2})$ and $\displaystyle 1s^22s^22p^3~(^4S_{3/2}) \to 1s^22s2p^3~(^5S_{2})$, to the $g-g$ (ground state to ground state) channel we obtain a better agreement. As can be seen from Table \ref{AdjustParam} ($g-g$ transition) and Table \ref{AdjustParam3} ($g-g$ + additional channels), the $A$ parameter increases by a factor 8.7 while the $B_p$s parameters vary by less than 2 \%,
    \item adding some channels to the $g-g$ one in C-like aluminum improves the agreement with the values given by Voronov,
    \item concerning Li- and Be-like aluminum, the cross sections involving the final excited states of He- and Li-like ions, respectively, are negligible, 
    \item the relative difference between our calculations and the data of Voronov (see Table \ref{ecart-voronov}) shows that the agreement increases with temperature.
\end{itemize}

Let us now compare our calculations with an experiment where the plasma is produced in a large bore theta-pinch \cite{Greve1982}. The experimental rate is obtained by a comparison of the time histories of spectral lines with calculated ones solving the set of coupled rate equations, and assuming that the upper level of the observed transition is populated only by collisional excitation from the ground state. Here, we have a specific temperature and density for each ion stage (see Table \ref{tableGreve}). It is worth mentioning that at these densities, the continuum lowering is very small compared to the ionization energy of the isolated ion (see our work, Ref. \cite{Benredjem2022}). The effect of the IPD on cross sections is then negligible. Table \ref{tableGreve} also contains quoted rates of Sampson and Golden \cite{Sampson1978}. These authors performed Coulomb-Born-exchange calculations \cite{Golden1977} including the inner-shell-autoionization contribution \cite{Sampson1979}. Their results are in good agreement with the measurements of Crandall \emph{et al.} \cite{Crandall1979} in the case of Li-like ions C$^{3+}$, N$^{4+}$ and O$^{5+}$ for energies from threshold to 1500 eV. However, as pointed out by Greve, the inner-shell autoionization contribution is found to be negligible for Al$^{10+}$ at a temperature around 220 eV. Tables \ref{tableGreve} and \ref{ecart-Greve-SG} show that 

\begin{itemize}
    \item our results present a good agreement with those of Sampson and Golden; the discrepancy is below 10 \%,
    \item the rates given by the Lotz formula agree with the experiment for the highest ion charges,
    \item our results show the best agreement with experiment for Al$^{10+}$ and Al$^{9+}$.
\end{itemize}

The difference between calculations and measurements are due in part to large experimental uncertainties, mentioned by the author (around 30 \% for Al$^{10+}$). Moreover, the ionization from metastable states is not well investigated. For instance, a large fraction of Al$^{9+}$ ions can occupy such a state, ($2p\,^3P$ level). Accounting for metastable states requires the knowledge of their population fractions. These can be estimated either by the Boltzmann law, if the environment of the ion is in thermal equilibrium, or by a collisional-radiative code. Moreover, the rate of the inner-shell ionization of Al$^{8+}$ ($1s^22s^22p\rightarrow 1s^22s2p+e^-$) is comparable with the one for ground-state ionization \cite{Kunze1971}.

\begin{table}[!ht]
\begin{center}
\begin{tabular}{|c|c|c|c|c|c|}
\hline
$k_{\rm B}T_e$ \;$\downarrow$ & Al$^{10+}$& Al$^{9+}$ & Al$^{8+}$ & Al$^{7+}$&Al$^{6+}$\\\hline
\multirow{3}{*}{30} & 2.76 10$^{-17}$ & 2.65 10$^{-16}$ & 4.09 10$^{-15}$ & 4.38 10$^{-14}$ & 3.31 10$^{-13}$\\
                    & 2.76 10$^{-17}$ & 2.65 10$^{-16}$ & 2.72 10$^{-15}$ & 2.54 10$^{-14}$ & 3.25 10$^{-14}$\\
                    & 1.99 10$^{-17}$ & 7.68 10$^{-17 }$ & 8.00 10$^{-16}$ & 3.50 10$^{-14}$ & 2.97 10$^{-13}$\\ \hline
\multirow{3}{*}{100} & 1.31 10$^{-12}$ & 4.95 10$^{-12}$ & 1.75 10$^{-11}$ & 5.79 10$^{-11}$ & 1.48 10$^{-10}$\\
                     & 1.31 10$^{-12}$ & 4.95 10$^{-12}$ & 1.05 10$^{-11}$ & 2.88 10$^{-11}$ & 1.37 10$^{-11}$\\
                     & 1.01 10$^{-12}$ & 2.56 10$^{-12}$ & 1.10 10$^{-11}$ & 4.45 10$^{-11}$ & 1.28 10$^{-10}$\\ \hline
\multirow{3}{*}{300} & 3.29 10$^{-11}$ & 1.04 10$^{-10}$ & 2.475 10$^{-10}$ & 5.15 10$^{-10}$ & 9.40 10$^{-10}$\\
                     & 3.29 10$^{-11}$ & 1.04 10$^{-10}$ & 1.42 10$^{-10}$ & 2.37 10$^{-10}$ & 8.62 10$^{-11}$\\
                     & 3.03 10$^{-11}$ & 7.40 10$^{-11}$& 2.00 10$^{-10}$ & 4.58 10$^{-10}$ & 8.97 10$^{-10}$\\ \hline
\multirow{3}{*}{1000} & 1.01 10$^{-10}$ & 3.18 10$^{-10}$ & 6.52 10$^{-10}$ & 1.06 10$^{-9}$ & 1.71 10$^{-9}$\\
                      & 1.01 10$^{-10}$ & 3.18 10$^{-10}$ & 3.77 10$^{-10}$ & 4.68 10$^{-10}$ & 1.56 10$^{-10}$\\
                      & 1.24 10$^{-10}$& 2.75 10$^{-10}$ & 5.84 10$^{-10}$ & 1.10 10$^{-9}$ & 1.87 10$^{-9}$\\ \hline
\end{tabular}
\end{center}
\caption{\label{Comparison1}Rate coefficient (in cm$^{3}$/s). The electron temperature $k_{\rm B}T_e$ is in eV. First entry: our result $g\rightarrow$ all final states, second entry: our result $g\rightarrow g$, third entry: Voronov's results, Ref. \cite{Voronov1997}.}
\end{table}

\begin{table}[!ht]
\begin{center}
\begin{tabular}{|c|r|r|r|r|r|r|}
\hline
&Al$^{10+}\hspace{.1cm}$&Al$^{9+}\hspace{.1cm}$&Al$^{8+}\hspace{.1cm}$&Al$^{7+}\hspace{.1cm}$&Al$^{6+}\hspace{.1cm}$\\
 \hline

$A$ & 1.3033 & 4.5440 & 10.0410 & 9.0895 & 20.4780 \\
$B_1$ & 1.5137 & 1.5595 & 1.2511 & 1.7965 & 1.5914 \\
$B_2$ & -1.8768 & -3.5505 & -3.1094 & -2.8386 & -2.8297 \\
$B_3$ & 0.9433 & 2.0352 & 1.4691 & 1.6552 & 1.6577 \\
\hline
\end{tabular}
\end{center}
\caption{\label{AdjustParam3}Values of the parameters when the cross section is $\sigma_2$. $A$ is in $10^{-19}$ cm$^2$ and the $B_p$s are dimensionless.}
\end{table}

\begin{table}[!ht]
\begin{center}
\begin{tabular}{|c|l|l|l|l|l|}\hline
$k_{\rm B}T_e$ \;$\downarrow$ & Al$^{10+}$& Al$^{9+}$ & Al$^{8+}$ & Al$^{7+}$&Al$^{6+}$\\\hline
30 & 32.9 & 132.0 & 182.0 & 22.5 & 10.8\\
100 & 26.1 & 67.1 & 46.8 & 26.4 & 14.5\\
300 & 8.2 & 28.8 & 19.2 & 11.1 & 4.6\\ 
1000 & -20.6 & 13.5 & 16.0 & -3.8 & -9.4\\\hline
\end{tabular}
\end{center}
\caption{\label{ecart-voronov} Relative difference $(X-Y)/\sqrt{X\,Y}\times 100$ (in \%), where $X$ represents our result and $Y$ that of Voronov \cite{Voronov1997} (see also Table \ref{Comparison1}).}
\end{table} 

\begin{table}
\begin{center}
\begin{tabular}{|l|l|l|l|l|l|}
\hline
 & Al$^{10+}$& Al$^{9+}$ & Al$^{8+}$ & Al$^{7+}$ & Al$^{6+}$\\
\hline
\multirow{2}{*}{$\begin{array}{l}k_{\rm B}T_e\\N_e\end{array}$} & 225 & 235 & 220 & 175 & 160 \\
& 3.2 10$^{16}$ & 2.7 10$^{16}$ & 2.1 10$^{16}$ & 1.5 10$^{16}$ & 1.3 10$^{16}$ \\\hline
\multirow{2}{*}{This work} & 1.90 10$^{-11}$ & 6.70 10$^{-11}$ & 1.50 10$^{-10}$ & 2.34 10$^{-10}$ & 4.16 10$^{-10}$ \\
& 1.90 10$^{-11}$ & 6.70 10$^{-11}$ & 8.64 10$^{-11}$ & 1.11 10$^{-10}$ & 3.83 10$^{-11}$ \\
\hline
Greve & 1.70 10$^{-11}$ & 5.80 10$^{-11}$ & 6.30 10$^{-11}$ & 1.70 10$^{-10}$ & 2.70 10$^{-10}$\\
\hline
Lotz & 2.29 10$^{-11}$ & 7.38 10$^{-11}$ & 6.19 10$^{-11}$ & 1.34 10$^{-10}$ & 2.95 10$^{-10}$\\
 \hline
SG & 2.10 10$^{-11}$ & 6.50 10$^{-11}$ & 1.50 10$^{-10}$ & 2.20 10$^{-10}$ & 4.30 10$^{-10}$\\
\hline
\end{tabular}
\end{center}
\caption{Rate coefficient (in cm$^3/$s). The electron temperature $k_{\rm B}T_e$ is in eV and the electron (number) density $N_e$ in cm$^{-3}$. This work: the first entry is our result involving $g\rightarrow$ all final states, the second entry is our result involving only $g\rightarrow g$. The last three rows contain the values of Greve \cite{Greve1982}, Lotz \cite{Lotz1968} and Sampson and Golden (SG) \cite{Sampson1979}.}
\label{tableGreve}
\end{table}

\begin{table}[!ht]
\begin{center}
\begin{tabular}{|c|l|l|l|l|l|}\hline
Relative differences (in \%) \;$\downarrow$ & Al$^{10+}$ & Al$^{9+}$ & Al$^{8+}$ & Al$^{7+}$ & Al$^{6+}$\\\hline
 & & & & &\\
$\displaystyle\frac{q_{2,\rm{B}}-q_{\mathrm{G}}}{\sqrt{q_{2,\rm{B}}\times q_{\mathrm{G}}}}\times 100$ & 11.1 & 14.4 & 89.5 & 32.1 & 43.6\\
 & & & & &\\
 \hline
 & & & & &\\
$\displaystyle\frac{q_{2,\rm{B}}-q_{\mathrm{SG}}}{\sqrt{q_{2,\rm{B}}\times q_{\mathrm{SG}}}}\times 100$ & -10.0 & 3.0 & 0.0 & 6.2 & -3.3\\
 & & & & &\\
 \hline
 & & & & &\\
$\displaystyle\frac{q_{\rm SG}-q_{\mathrm{G}}}{\sqrt{q_{\rm SG}\times q_{\mathrm{G}}}}\times 100$ & 21.2 & 11.4 & 89.5 & 25.9 & 47.0\\
 & & & & &\\
 \hline
 & & & & &\\ 
 $\displaystyle\frac{q_{\rm{L}}-q_{\mathrm{G}}}{\sqrt{q_{\rm{L}}\times q_{\mathrm{G}}}}\times 100$ & 29.9 & 24.1 & -1.76 & -23.9 & 8.86\\
  & & & & &\\
 \hline
\end{tabular}
\end{center}
\caption{\label{ecart-Greve-SG} Relative differences, $(X-Y)/\sqrt{X\,Y}\times 100$ (in \%), between measurements and calculations. $q_{2,\rm{B}}$: our calculations, $q_{\rm G}$: measurements, $q_{\rm L}$: Lotz, $q_{\rm SG}$: Sampson and Golden. See also Table \ref{tableGreve}.}
\end{table} 

\clearpage

\section{Conclusion and prospective}\label{sec8}

We proposed a new semi-empirical formula for the calculation of the electron-impact ionization cross sections. Four adjustable parameters are necessary to achieve a good fit of measured or calculated cross sections. In the latter case, we relied on the cross sections calculated with the FAC code. In contrast to Kim's formula, ours satisfies the threshold condition, \textit{i.e.}, $\sigma_2=0$ at $E=E_i$. Moreover, our formula is more reliable and robust when interpolating or extrapolating. The plasma density effect known as the ionization potential depression may affect the cross section values. In this study, we used the formula of Ecker and Kr\"{o}ll to estimate the variation of the ionization energy, and we showed that the continuum lowering has an important effect on ionization by free-electron impacts. In the case of aluminum ions, the adjustable parameters were determined by a fit of calculations with the FAC code. Alternatively, it would be interesting to determine the parameters of $\sigma_2$ from the time-dependent close-coupling cross section \cite{Badnell1998,Pindzola2021}, for example in Al$^{+2}$, and check whether it improves the agreement with the experimental results \cite{Crandall1982}.

In a second part, we calculated the rate coefficients analytically, at the Boltzmann limit, and numerically, in the case of a Fermi-Dirac distribution. Our results for aluminum ions were then compared with measurements and other calculations, in a large temperature range. We discussed the discrepancies, when they occurred, and suggested to account for additional processes, such as the ionization from metastable states (for example in Be-like aluminum) or from inner-shell (for example in B-like aluminum). 

There is a strong need to perform new experiments where all the channels participating to the measured rates can be clearly identified. Otherwise, any comparison with calculations will not be meaningful. Amongst the channels, the ionization from excited states, such as metastable states, plays a role. To account for these channels in rate calculations, the level-population fractions must first be calculated.


\providecommand{\newblock}{}

\end{document}